\documentclass[twocolumn,showpacs,prb,amsmath,amssymb,floatfix]{revtex4}
\usepackage{graphicx}
\usepackage{amsmath}
\usepackage{amsbsy}
\usepackage{amsfonts}

\begin{document}

\author{J. Planelles}
\email{josep.planelles@uji.es}
\author{J. I. Climente}
\affiliation{Departament de Quimica Fisica i Analitica, Universitat Jaume I, Box 224, E-12080 Castell\'o, Spain}
\date{\today}

\title{Magnetic field implementation in multiband k$\cdot$p Hamiltonians of holes in semiconductor heterostructures}

\begin{abstract} 

We propose an implementation of external homogeneous magnetic fields in k$\cdot$p Hamiltonians for holes in 
heterostructures, in which we made use of the minimal coupling prior to introduce the envelope function approximation. 
Illustrative calculations for holes in InGaAs quantum dot molecules show that the proposed Hamiltonian  outperforms
standard Luttinger model [Physical Review {\bf 102}, 1030 (1956)] describing the experimentally observed magnetic response.
The present implementation culminates our previous proposal [Phys. Rev. B {\bf 82}, 155307 (2010)].
\end{abstract}


\maketitle
 
\section{Introduction}

The interaction of a magnetic field with a charged particle with spin comes into the Hamiltonian 
through coupling to the total (orbital plus spin) angular momentum.
In crystals, the total angular momentum is the sum of the Bloch angular momentum, which contains 
atomic orbital and spin contributions, and the envelope orbital angular momentum.\cite{Cardona_book,Loher_book}
Determining the coupling constant (g factor) is an important requirement to
study the magnetic properties of materials.
In bulk systems, the value of the g factor is strongly influenced by band mixing, 
spin-orbit interaction and crystal anisotropy.\cite{roth} 

In the last years, there is increasing interest in controlling and exploiting the g factor of 
carriers confined in semiconductor quantum dots (QDs) for spin preparation, conservation and
manipulation aiming at spintronic and quantum information devices.\cite{Awschalom_book,Doty_Phys,ChenPRB,EconomouPRB,KhaetskiiPRB,DeaconPRB}
The magnetism of these structures is significantly different from that of bulk crystals
because the weak magnetic confinement is supplemented by a strong spatial confinement.
The latter has a profound effect on the energy structure, carrier-carrier interactions,
band mixing and spin-orbit interactions,\cite{Pawel_book} which, in turn, influence
the g factor value. As a matter of fact, it has been shown that quantum confinement in
QDs leads to strongly anisotropic g factors for both electrons and holes,\cite{PradoPRB,BayerPRB,MayerAlegrePRL}
as well as to a quenching of the g factor value.\cite{pryor06,KimPRB,RoddaroPRL} 

Up to date, most theoretical studies investigating the magnetic response of QDs
rely on effective mass and $\mathbf k \cdot \mathbf p$ models. 
The standard inclusion of magnetic fields in $\mathbf k \cdot \mathbf p$ Hamiltonians consists in 
replacing the canonical momentum $\mathbf p$ by the kinetic momentum  $\mathbf p - q  \mathbf A$,
supplemented with the spin Zeeman term, in the differential equation fulfilled by the envelope function
(here $q$ is the carrier charge and $\mathbf A$ the potential vector defining the magnetic field).\cite{Luttinger,Voon_book,MlinarPRB}
This approximation, hereafter referred to as the Luttinger approximation, has been successful in 
explaining several experimental observations in heterostructures,\cite{Pawel_book,Chakra_book,LeePE} 
it is implemented in the widely employed semiconductor software package \emph{nextnano},\cite{BirnerIEE,AndlauerPRB}
and it is currently being used to determine the g factors of confined carriers.\cite{pryor06,AndlauerPRB,bree}
However, a number of situations have been identified where it provides qualitatively incorrect predictions.
For example, in quantum rings under axial magnetic fields, the Luttinger approximation predicts the optical 
gap to decrease with the field strength,\cite{ClimentePRB} contrary to magneto-photoluminesence observations.\cite{HaftPE}
Similarly, in vertically coupled QDs, the Luttinger approximation predicts that an axial magnetic field
can tune the hole tunneling rate,\cite{ClimenteAPL} but this effect is not observed in related experiments.\cite{PlanellesPRB}
The underlying reason is that the Luttinger approximation includes off-diagonal magnetic terms in the
Hamiltonian, which artificially enhance the heavy hole-light hole (HH-LH) band mixing.\cite{ClimentePRB,PlanellesPRB}

In Ref.~\onlinecite{JPCM}, an alternative formulation of the magnetic interaction was suggested,
in which the replacement of the canonical  momentum by the kinetic one is carried out prior to the
envelope function approximation (EFA). The resulting Hamiltonian has no off-diagonal magnetic terms directly coupling
HH and LH, and the results become then consistent with the experimental measurements.\cite{ClimentePRB,PlanellesPRB}

In the present work, we extend Ref.~\onlinecite{JPCM} theory in order to account for the spin Zeeman term,
and identify the coefficients that should accompany the magnetic terms in this approximation, 
which were pending clarification.\cite{PlanellesPRB} The remote band influence is considered through
the zero-field effective masses, which are known to provide a meaningful description even in strongly
confined QDs\cite{norris,wlodek1,pokatilov,wlodek2}, 
and effective $g$-factors.
We run calculations comparing this model with the Luttinger approximation and show that 
the magnitude of the Zeeman splitting we estimate is closer to experimental values for vertically coupled InGaAs QDs.
Thus, our theory offers improved accuracy in current attempts to understand and predict g factor values of holes in QDs.

The paper is organized as follows. In Section \ref{s:theo}, we derive the multiband k$\cdot$p-EFA Hamiltonian
for holes in the presence of a magnetic field. Starting from a general formulation, the Hamiltonian is then
particularized to the case of QDs with axial symmetry. In Section \ref{s:res}, we use the obtained Hamiltonian 
to calculate the Zeeman splitting of vertically coupled QD molecules. The results are compared with previous
implementations of the magnetic field and experimental data. Conclusions are given in Section \ref{s:conc}. 

\section{Theory}\label{s:theo}
\subsection{The Hamiltonian}
The classical Hamiltonian of a charged particle with anisotropic mass response to 
 external forces, subject to the action of a magnetic field defined by a potential vector $\mathbf A$ is:
\begin{equation}
\label{eq1}
{\cal H}=\sum_i \frac{\pi_i^2}{2 m_i}= \sum_i \frac{(p_i-q\, A_i)^2}{2 m_i},
\end{equation}
\noindent where $q$ is the charge and $m_i$, $i=x,y,z$ the anisotropic mass.\\

\noindent Elemental particles, in addition to charge, have spin. We can introduce heuristically spin  by making the formal replacement
$\boldsymbol {\pi} \to \boldsymbol {\sigma} \cdot \boldsymbol {\pi}$ in the above equation and taking into account the next two 
identities involving vectorial operators:

\begin{eqnarray}
\label{eq2}
(\boldsymbol {\sigma} \cdot \mathbf a) (\boldsymbol {\sigma} \cdot \mathbf b) &=& \mathbf a \cdot \mathbf b +i \boldsymbol {\sigma}\cdot (\mathbf a \times \mathbf b)\\
\nonumber \\
\label{eq3}
\mathbf p \times \mathbf A+ \mathbf A \times \mathbf p &=& -i \hbar \nabla \times \mathbf A = -i \hbar \mathbf B
\end{eqnarray}
\noindent In the above equations $\mathbf a$, $\mathbf b$ are vector operators, the components of $\boldsymbol {\sigma}$ are the Pauli matrices and $\mathbf B$ represents the magnetic field. This formal replacement
turns the kinetic energy $T=\frac{(\mathbf p - q \, \mathbf A)^2}{2m}$ into 
\begin{equation}
\label{eq3b}
T_D=\frac{(\mathbf p - q \, \mathbf A)^2}{2m}-\frac{q \hbar}{2m}\, \boldsymbol {\sigma}\cdot \mathbf B,
\end{equation}
\noindent as it should appear in the Dirac equation. If the mass is anisotropic  we should write instead:
\begin{equation}\begin{split}
\label{eq4}
T_D & =\frac{1}{2} \left( \sum_i \frac{\sigma_i \pi_i}{m_i}\right)\left( \sum_j \frac{\sigma_j \pi_j}{m_j}\right) =  \\
    & =\frac{1}{2} (\boldsymbol {\sigma}\cdot  \overline{\boldsymbol \pi}) (\boldsymbol {\sigma}\cdot \overline{ \boldsymbol \pi})  
   = \frac{1}{2} \overline{\pi}^2 + \frac{i}{2} \boldsymbol {\sigma}\cdot \overline{\boldsymbol\pi}\times \overline{ \boldsymbol \pi}
\end{split} \end{equation}
\noindent where $\overline{\pi}_i=\frac{\pi_i}{\sqrt{m_i}}$.\\

\noindent The $\boldsymbol {\sigma}\cdot \overline{\boldsymbol \pi}\times \overline{ \boldsymbol \pi}$ term in (\ref{eq4}) can be expanded as follows:
\begin{equation} \label{eq5}
\begin{split}
\boldsymbol {\sigma}\cdot \overline{\boldsymbol \pi}\times \overline{ \boldsymbol \pi} & = \sigma_x \frac{1}{\sqrt{m_y m_z}} [\pi_y,\pi_z]+
\sigma_y \frac{1}{\sqrt{m_x m_z}} [\pi_z,\pi_x] \\
 & +\sigma_z \frac{1}{\sqrt{m_x m_y}} [\pi_x,\pi_y]
\end{split}
\end{equation}
\noindent where $[\pi_i,\pi_j]= \pi_i \pi_j- \pi_j \pi_i$.\\

\noindent We may define $\overline{\boldsymbol \sigma}$ with components $\frac{\sigma_i}{\sqrt{m_j\, m_k}}$, where $i,j,k$ represent a cyclic  permutations of $x,y,x$. Then, we have:

\begin{equation}
\label{eq6}
\boldsymbol {\sigma}\cdot \overline{\boldsymbol \pi}\times \overline{ \boldsymbol \pi} =  \overline{\boldsymbol \sigma}\cdot (\boldsymbol \pi 
\times \boldsymbol \pi).
\end{equation}

\noindent In the other hand, since $\boldsymbol \pi \times \boldsymbol \pi =(\mathbf p -q \mathbf A)\times (\mathbf p -q \mathbf A) = 
-q (\mathbf p \times \mathbf A + \mathbf A \times \mathbf p) = i \, q \, \hbar \nabla \times A=i \, q \, \hbar \, \mathbf B$ we find out
that,
\begin{equation}
\label{eq7}
\frac{i}{2}\boldsymbol {\sigma}\cdot \overline{\boldsymbol \pi}\times \overline{ \boldsymbol \pi} 
         = -\frac{q \hbar}{2} \, \overline{\boldsymbol {\sigma}} \cdot \mathbf B.
\end{equation}

\noindent From now on, we will restrict ourselves to the case of axial symmetry, i.e., to the particular case $\mathbf B= B_0 \, \mathbf k$, $m_x=m_y=m_{\perp}$. Then, the above term turns into $-\frac{q \hbar}{2 m_{\perp}} \sigma_z \, B_0$ and the complete Hamiltonian reads,

\begin{equation}
\label{eq8}
\begin{split}
{\cal H} &= \sum_{\alpha=\perp,z}\, (p_{\alpha}-q\, A_{\alpha}) \frac{1}{2 m_{\alpha}} (p_{\alpha}-q\, A_{\alpha}) \\
 & -\frac{q \hbar}{2 m_{\perp}} \sigma_z \, B_0 = {\cal T}-\frac{q \hbar}{2 m_{\perp}} \sigma_z \, B_0 
\end{split}
\end{equation}

\noindent By employing the symmetric gauge $\mathbf A=\frac{B_0}{2}\; [-y,x,0]$, we see that $A_z=0$. Then,
$z$ component of the kinetic energy is not affected, ${\cal T}_z=p_z \, \frac{1}{2 m_z} \, p_z$ , while {\it in-plane} component is:
\begin{equation}
\label{eq9}
{\cal T_{\perp}}= p_{\perp} \, \frac{1}{2 m_{\perp}} \, p_{\perp}+ \frac{q^2 A_{\perp}^2}{2 m_{\perp}}- 
                    \frac{q A_{\perp}}{2 m_{\perp}}\cdot p_{\perp}- \frac{q}{2} p_{\perp} \cdot \frac{A_{\perp}}{m_{\perp}}
\end{equation}

\noindent Since $m_{\perp}(\rho, z)$ then, $p_{\perp} (\frac{1}{m_{\perp}})=-\frac{i \, \hbar}{\rho} [x,y,0] \, \frac{\partial }{\partial \rho} (\frac{1}{m_{\perp}})$ and $p_{\perp} \cdot \frac{A_{\perp}}{m_{\perp}} \, \Psi= \frac{A_{\perp}}{m_{\perp}}\cdot p_{\perp} \, \Psi$, so that
the {\it in-plane} component of the kinetic energy results,
\begin{equation}\label{eq9b}
\begin{split}
{\cal T_{\perp}}&=  p_{\perp} \, \frac{1}{2 m_{\perp}} \, p_{\perp}+ \frac{q^2 A_{\perp}^2}{2 m_{\perp}}- 
                    \frac{q }{m_{\perp}} \; A_{\perp}\cdot p_{\perp} \\ 
				&= -\hbar^2 \nabla_{\perp} \, \frac{1}{2 m_{\perp}} \, \nabla_{\perp}+ \frac{q^2 B_0^2 \rho^2}{8 m_{\perp}}- 
                    \frac{q \, B_0}{2 m_{\perp}} \; (x \,\hat p_y -y \hat p_x)	
\end{split}
\end{equation}

\noindent and the complete Hamiltonian is in turn 
\begin{equation}\label{eq10}
{\cal H}= -\sum_{\alpha=\perp,z} \nabla_{\alpha} \, \frac{1}{2 m_{\alpha}} \, \nabla_{\alpha} +\frac{q^2 B_0^2 \rho^2}{8 m_{\perp}}- 
                    \frac{q \, B_0}{2 m_{\perp}} \, \hat L_z -\frac{q \, B_0}{2 m_{\perp}} \, \sigma_z
\end{equation}

\noindent This equation particularized to effective electrons ($m > 0$, $q=-1$) is, 
\begin{equation}
\label{eq11}
{\cal H}_e= -\sum_{\alpha=\perp,z} \nabla_{\alpha} \, \frac{1}{2 m_{\alpha}} \, \nabla_{\alpha}  +\frac{B_0^2 \rho^2}{8 m_{\perp}}+ 
                    \frac{B_0}{2 m_{\perp}} \, (\hat L_z + \sigma_z)
\end{equation}
 
\noindent What about holes ($m=-|m| < 0$, $q=1$)? Holes are tricky particles that require a careful tackle. 
We know that in the case of a one-band model, electrons and holes energy dispersion are mirror image of each other. Then, we should assume that,

\begin{equation}
\label{eq12}
{\cal H}_h= \sum_{\alpha=\perp,z} \nabla_{\alpha} \, \frac{1}{2 |m_{\alpha}|} \, \nabla_{\alpha} - \frac{B_0^2 \rho^2}{8 |m_{\perp}|}- \frac{B_0}{2 |m_{\perp}|} \, (\hat L_z + \sigma_z)
\end{equation}

\subsection{Envelope and Bloch functions}

In solid state physics the  wave function $|\Psi (\mathbf{r}) \rangle$ is expressed as
a sum of products $|\Psi(\mathbf{r})\rangle=\sum_i^N{|J J_z\rangle_i |f\rangle_i}$, where $|J J_z\rangle$ are the Bloch band-edge  
and $|f\rangle$ are the envelope functions.\cite{Loher_book,Voon_book} Let us call ${\cal H}^0$ to the Hamiltonian, eq. (\ref{eq12}),
in absence of magnetic field and ${\cal H}^{(B)}$ to the second and third terms of this equation describing the action of the
external magnetic field. The Luttinger-Kohn ${\mathbb H}^0$  matrix Hamiltonian operator\cite{Luttinger} acting on the envelope vector function
can be obtained by applying ${\cal H}^0$ onto $|\Psi (\mathbf{r}) \rangle$. Then, left-multiplying by the different Bloch functions $\langle J J_z|_i$
and integrating over the unit cell. Afterwards, the effect of remote bands is incorporated by replacing the actual mass by effective masses
in the matrix elements of  ${\mathbb H}^0$.\\

\noindent In the presence of magnetic field  ${\mathbb H}^0$ must be supplemented by ${\mathbb H}^{(B)}$ coming from ${\cal H}^{(B)}$ and 
$|\Psi (\mathbf{r}) \rangle$, through a similar procedure. Since all terms in ${\cal H}^{(B)}$ act as pure multiplicative operators on
the envelope function components except for $\hat L_z$, we have that,\cite{JPCM}

\begin{equation}
\label{ebf1}
{\cal H}^{(B)} |J J_z\rangle |f\rangle = |f\rangle {\cal H}^{(B)} |J J_z\rangle - |J J_z\rangle \frac{B_0}{2|m_{\perp}|}\, \hat L_z |f\rangle
\end{equation}
\noindent Axially symmetric systems have well defined z-component $F_z$ of the total angular momentum and, additionally, the components
of the envelope function associated to the Bloch function $|J J_z\rangle$ have also a well defined $M=(F_z-J_z)$ orbital angular momentum,\cite{efros,vahala} so that
$\hat L_z |f\rangle=(F_z-J_z)|f\rangle$. Then, we calculate the $(J' J_z',J J_z)$ matrix element of ${\mathbb H}^{(B)}$ as follows:

\begin{multline}\label{ebf2}
\langle J' J_z'| {\cal H}^{(B)} \left(|JM\rangle |f\rangle\right) =  \\
\langle J' J_z'|\left( {\cal H}^{(B)} -\frac{F_z-J_z}{2|m_{\perp}|} B_0 \right) |J J_z\rangle \cdot |f\rangle
\end{multline}
\noindent i.e.,
\begin{equation}
\label{ebf3}
{\mathbb H}^{(B)}=-\frac{B_0^2 \rho^2}{8 |m_{\perp}|} {\mathbb I}-\frac{F_z-J_z}{2 |m_{\perp}|} B_0 \; {\mathbb I}-\frac{B_0}{2 |m_{\perp}|}\left( {\mathbb L}_z + \sigma_z\right)
\end{equation}

\noindent Next, we incorporate de effect of the remote bands like in electrons: the mass $m_{\perp}$ arising in the
two first terms of ${\mathbb H}^{(B)}$ is replaced by the effective
mass parameter appearing in the corresponding matrix elements of ${\mathbb H}^0$, while the third term in
${\mathbb H}^{(B)}$ becomes $-\kappa \mu_B B_0 {\mathbb J}_z$. The procedure yields the following non-zero matrix elements for the 6x6 valence Hamiltonian for zinc-blende crystals, which includes heavy hole, light hole and split-off bands. 

\small
\begin{eqnarray}
\label{ebf4}
\nonumber
{\mathbb H}^{(B)}_{11}&=&  -(\gamma_1+\gamma_2)\left[\frac{B_0^2 \rho^2}{8}+\frac{B_0}{2} (F_z-3/2)\right] - \frac{3}{2} \kappa \mu_B B_0  \\
\nonumber
{\mathbb H}^{(B)}_{22}&=&  -(\gamma_1-\gamma_2)\left[\frac{B_0^2 \rho^2}{8}+\frac{B_0}{2} (F_z-1/2)\right] - \frac{1}{2} \kappa \mu_B B_0 \\	
\nonumber
{\mathbb H}^{(B)}_{33}&=&  -(\gamma_1-\gamma_2)\left[\frac{B_0^2 \rho^2}{8}+\frac{B_0}{2} (F_z+1/2) \right] + \frac{1}{2} \kappa \mu_B B_0\\
\nonumber
{\mathbb H}^{(B)}_{44}&=&  -(\gamma_1+\gamma_2)\left[\frac{B_0^2 \rho^2}{8}+\frac{B_0}{2} (F_z+3/2) \right] + \frac{3}{2} \kappa \mu_B B_0 \\
\nonumber
{\mathbb H}^{(B)}_{55}&=&  -\gamma_1 \left[\frac{B_0^2 \rho^2}{8}+\frac{B_0}{2} (F_z-1/2) \right] - \frac{1}{2} \kappa' \mu_B B_0 \\
{\mathbb H}^{(B)}_{66}&=&  -\gamma_1 \left[\frac{B_0^2 \rho^2}{8}+\frac{B_0}{2} (F_z+1/2) \right] + \frac{1}{2} \kappa' \mu_B B_0 
\end{eqnarray}
\normalsize

\noindent where $\mu_B=|q|/2m$ is the Bohr magneton, while $\kappa$ and $\kappa'$ are effective g factors for 
holes which become $4/3$ and $2/3$ respectively if we remove the contribution of the remote bands. 

Note that the magnetic terms in Eq.~(\ref{ebf4}) differ  from those of the Luttinger approximation 
(see Table 3 in Ref.~\onlinecite{JPCM}). In particular, there are no off-diagonal magnetic terms mixing HH and LH
subbands. They also differ from our previous proposal\cite{JPCM}
in two aspects: (i) the spin degree of freedom is now included, and (ii) the remote bands contribution to the 
linear-in-$B$ term coming from the Bloch function (third term in Eq.~(\ref{ebf3})) is now 
included through hole g factors ($\kappa,\kappa'$).

\section{Illustrative calculations}\label{s:res}

In this section we implement the magnetic terms described above in a 4-band
k$\cdot$p Hamiltonian coupling HH and LH states. Hereafter we refer to this
as ${\cal H}_{\kappa}$. For comparison, we also implement the magnetic terms 
following the Luttinger approximation, ${\cal H}_{lutt}$ (Table 3 of Ref.~\onlinecite{JPCM} 
but adding diagonal spin terms, $-\kappa \mu_B B_0 {\mathbb J}_z$),
and our previous proposal
${\cal H}_{mass}$ (four-band Hamiltonian taken from Ref.~\onlinecite{carlos}).

To test the performance of the different Hamiltonians, we compare with available 
experimental data for an InGaAs/GaAs QD molecule subject to a longitudinal magnetic 
field.\cite{PlanellesPRB}
The Hamiltonian is solved numerically for a structure formed by two vertically 
stacked cylindrical QDs. The QDs have radius $R=15$ nm and height $H=2$ nm,
with an interdot barrier of thickness $S=2.8$ nm. InGaAs Luttinger parameters
are used for the masses, ($\gamma_1=11.01$, $\gamma_2=4.18$, and $\gamma_3=4.84$),\cite{VurgaftmanJAP}
and a valence band offset of $0.2$ eV is taken at the interfaces.
In bulk, the $\kappa$ constant takes a value of $7.68$ (for pure InAs).\cite{LawaetzPRB}
However, quantum confinement severely quenches this value. In QDs, one can 
safely disregard the contribution from remote bands, and simply consider
that coming from the HH-LH subband coupling.\cite{bree} 
Therefore, we take $\kappa=4/3$.

\begin{figure}[h]
\includegraphics[width=0.5\textwidth]{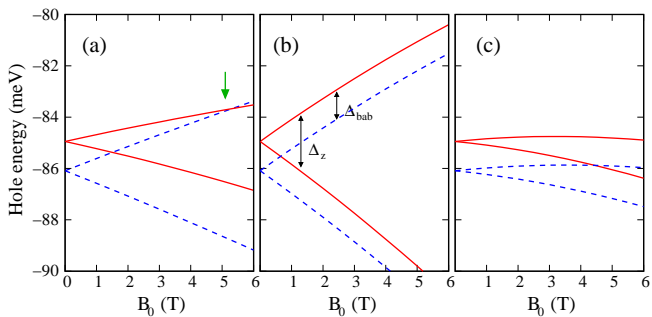}
\caption{Magnetic field dispersion of the highest hole states in a QD molecule
calculated with different implementations of the magnetic terms.
(a): ${\cal H}_{lutt}$, 
(b): ${\cal H}_{mass}$, 
(c): ${\cal H}_{\kappa}$.
Solid and dashed lines are used for the bonding and antibonding hole states.
The arrow in (a) indicates the bonding-antibonding ground state reversal.}
\label{fig:Bdisp}
\end{figure}

Figure \ref{fig:Bdisp} shows the energy of the highest valence band states.
These are the $|Fz|=3/2$ hole states with bonding (solid lines) 
and antibonding (dashed lines) molecular character.\cite{ClimentePRB2}
Panels (a), (b) and (c) correspond to estimates obtained with  ${\cal H}_{lutt}$,
${\cal H}_{mass}$, and ${\cal H}_{\kappa}$, respectively.
One can see there are conspicuous differences in the energy spectra.
For example, both ${\cal H}_{lutt}$ and ${\cal H}_{mass}$ predict that for the
ground state, the $B$-linear term dominates over the $B$-quadratic (diamagnetic) one,
so that its energy increases with the field.
This would imply a decrease of the excitonic gap, in sharp contrast with photoluminescence 
experiments of InGaAs QDs, where the gap increases quadratically, 
indicating that the diamagnetic term is dominant.
This is precisely the situation predicted by ${\cal H}_{\kappa}$, Fig.~\ref{fig:Bdisp}(c).

\begin{figure}[h]
\includegraphics[width=0.4\textwidth]{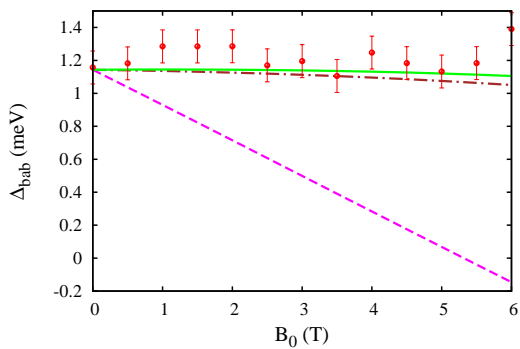}
\caption{Bonding-antibonding energy splitting as a function of the magnetic field.
Dashed line: ${\cal H}_{lutt}$.
Dashed-dotted line: ${\cal H}_{mass}$.
Solid line: ${\cal H}_{\kappa}$.}
\label{fig:bab}
\end{figure}

Further insight is obtained by comparing energy differences within each spectrum.
We first compare the energy splitting between the bonding and antibonding states, 
$\Delta_{bab}$ as a function of the magnetic field. 
Fig.~\ref{fig:bab} shows $\Delta_{bab}$ calculated with the three Hamiltonians.
${\cal H}_{lutt}$ (dashed line) predicts that the energy splitting decreases with $B_0$,
becomes zero at $B_0=5.3$ T and negative afterwards, which 
means that the ground state has changed from bonding to antibonding character.
This ground state crossing is indicated by a green arrow in Fig.~\ref{fig:Bdisp}(a).
The modulation of $\Delta_{bab}$ with longitudinal magnetic fields is a consequence 
of the off-diagonal magnetic terms in ${\cal H}_{lutt}$.\cite{ClimenteAPL} 
However, no such behavior is found in experiments, where $\Delta_{bab}$ remains 
roughly constant with the field. This is shown by the symbols in Fig.~\ref{fig:bab},
which represent experimental data taken from Ref.~\onlinecite{PlanellesPRB}.
Clearly, both ${\cal H}_{mass}$ (dashed-dotted line) and ${\cal H}_{\kappa}$ (solid line)
succeed in reproducing the approximately constant value of $\Delta_{bab}$. 

\begin{figure}[h]
\includegraphics[width=0.4\textwidth]{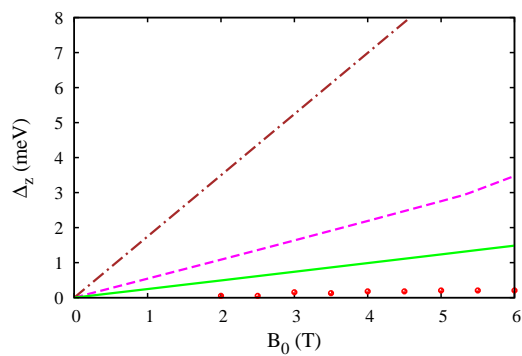}
\caption{Zeeman splitting as a function of the magnetic field.
Dashed line: ${\cal H}_{lutt}$.
Dashed-dotted line: ${\cal H}_{mass}$.
Solid line: ${\cal H}_{\kappa}$.}
\label{fig:zeeman}
\end{figure}

In order to discriminate between ${\cal H}_{mass}$ and ${\cal H}_{\kappa}$,
in Figure \ref{fig:zeeman} we compare the Zeeman splitting of the ground, 
$\Delta_{z}$, calculated with all three Hamiltonians and the experimental
values, represented by symbols. It can be seen that ${\cal H}_{mass}$ 
(dashed-dotted line) vastly overestimates the Zeeman splitting, while 
${\cal H}_{\kappa}$ (solid line) offers the closest description. 
It is worth noting that the experimental values of $\Delta_z$ are even 
smaller than those predicted by ${\cal H}_{\kappa}$. The inclusion of
strain and piezoelectric effects may be relevant for a quantitatively
improved description.\cite{bree,AndlauerPRB2}

\begin{figure}[h]
\includegraphics[width=0.4\textwidth]{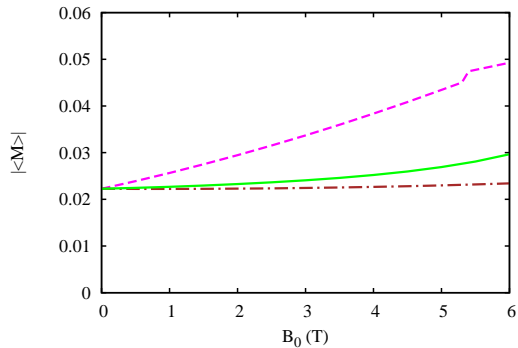}
\caption{Envelope angular momentum expectation value of the ground state
as a function of the magnetic field.
Dashed line: ${\cal H}_{lutt}$.
Dashed-dotted line: ${\cal H}_{mass}$.
Solid line: ${\cal H}_{\kappa}$.}
\label{fig:MaZinger}
\end{figure}

Last, we compare the envelope angular momentum admixture obtained
with the different Hamiltonians. 
In a four-band model, the hole states of cylindrical QDs are 
four-component spinors of the form:

\begin{equation}
|F_z,n \rangle = \sum_{J_z=-3/2,3/2}\,|f_{M}\rangle \, |\frac{3}{2} J_z \rangle.
\end{equation}

\noindent where $n$ is the main quantum number and $M=F_z-J_z$ is the envelope 
azimuthal angular momentum of a given component. We calculate the expectation 
value of the ground state envelope angular momentum at different fields and 
plot the results in Figure \ref{fig:MaZinger}. 
At zero field $\langle M \rangle =0.02$, indicating that the ground state
largest component is by far the HH ($|J_z|=3/2$) with $M=0$, with a small
admixture with finite $M$ components. When the magnetic field is switched
on, ${\cal H}_{lutt}$ predicts a faster increase of $\langle M \rangle$
than ${\cal H}_{mass}$ or ${\cal H}_{\kappa}$ (the bump at $B_0=5.4 T$
is due to the bonding-antibonding reversal).
Since the degree of envelope angular momentum admixture is critical in 
determining the effective g factor of holes,\cite{bree}
and we have shown that only ${\cal H}_{\kappa}$ provides a consistent
description of the magnetic response, Fig.~\ref{fig:MaZinger} implies
that the widely used Luttinger approximation is likely to overestimate 
the g factor values in confined systems.

\section{Summary}\label{s:conc}

We have derived a multiband k$\cdot$p Hamiltonian for valence holes
confined in heterostructures subject to an external magnetic field.
The magnetic field has been implemented 
incorporating the minimal coupling ${\mathbf p} \to {\mathbf p}-q {\mathbf A}$
prior to introduce the EFA. 
The inclusion of the remote bands has been considered through effective masses 
for the envelope function terms and effective g factors for the magnetic terms 
originating in the (unit cell) Bloch functions. 
For QDs, owing to the strong confinement, the latter have been replaced by 
the bare hole values, disregarding the influence of remote bands.

The resulting Hamiltonian has been compared with the widely employed
Luttinger approximation and our previous proposal. 
When tested against experimental data for InGaAs QDs under axial magnetic fields, 
the Hamiltonian presented in this work clearly outperforms the others.
In particular, it succeeds in simultaneously describing the increase of the 
excitonic gap with the field, the constant splitting between bonding and 
antibonding states and the small Zeeman splitting observed in photoluminescence
experiments, with no fitting parameters.

The Hamiltonian we have formulated is expected to improve current attempts 
to estimate the g factors of holes confined in QDs for spintronic, quantum
information and optical applications.\\

\begin{acknowledgments}
We thank M. Doty, A. S. Bracker and D. Gammon for sharing experimental data.
Support from MICINN project CTQ2011-27324, UJI-Bancaixa project P1·1B2011-01 
and the Ramon y Cajal program (JIC) is acknowledged.\\
\end{acknowledgments}

\end{document}